\documentclass{optica-article}

\journal{opticajournal} 

\articletype{Research Article}

\usepackage{graphicx}
\usepackage{dcolumn}
\usepackage{bm}
\usepackage{hyperref}

\usepackage[bb=boondox]{mathalfa}
\usepackage{braket}
\usepackage{dsfont}

\usepackage[colorinlistoftodos,prependcaption,textsize=tiny]{todonotes} 

\usepackage{xcolor}

\DeclareMathOperator{\tvd}{TVD}

\DeclareMathOperator{\tmsv}{TMSV}

\begin{document}

\title{Boosting Photon-Number-Resolved Detection Rates of Transition-Edge Sensors by Machine Learning}

\author{
Zhenghao Li,\authormark{1, $\dagger$,\#} 
Matthew J.H. Kendall,\authormark{1, $\dagger$}
Gerard J. Machado, \authormark{1}
Ruidi Zhu, \authormark{1, 2}
Ewan Mer, \authormark{1}
Hao Zhan, \authormark{1, 3}
Aonan Zhang, \authormark{1}
Shang Yu, \authormark{1} 
Ian A. Walmsley, \authormark{1} 
and Raj B. Patel \authormark{1, *}
}

\address{
\authormark{1}Department of Physics, Imperial College London, Prince Consort Road, London, SW7 2AZ, UK\\
\authormark{2}Department of Applied Physics, Yale University, 15 Prospect Street, New Haven, Connecticut 06511, USA\\
\authormark{3}College of Engineering and Applied Sciences, Nanjing University, 163 Xianlin Road, Nanjing 210093, China
}

\email{\authormark{$\dagger$} These authors contributed equally to this work. \\
\authormark{\#} zhenghao.li21@imperial.ac.uk\\
\authormark{*} raj.patel1@imperial.ac.uk} 


\begin{abstract*} 
Transition-Edge Sensors (TESs) are very effective photon-number-resolving (PNR) detectors that have enabled many photonic quantum technologies. However, their relatively slow thermal recovery time severely limits their operation rate in experimental scenarios compared to leading non-PNR detectors. In this work, we develop an algorithmic approach that enables TESs to detect and accurately classify photon pulses without waiting for a full recovery time between detection events. We propose two machine-learning-based signal processing methods: one supervised learning method and one unsupervised clustering method. By benchmarking against data obtained using coherent states and squeezed states, we show that the methods extend the TES operation rate to 800~kHz, achieving at least a four-fold improvement, whilst maintaining accurate photon-number assignment up to at least five photons. Our algorithms will find utility in applications where high rates of PNR detection are required and in technologies which demand fast active feed-forward of PNR detection outcomes. 

\end{abstract*}

\section{Introduction}
Transition-Edge Sensors (TESs) are photon-number resolving (PNR) detectors that represent a pinnacle of precision in modern optical detection technology.
These superconducting sensors operate at the threshold between superconducting and normal conductive states, leveraging this transition phase to act as a micro-calorimeter that can measure the minute temperature changes induced by photon absorption~\cite{lita_counting_2008, lita_superconducting_2010}. Their photon-number resolving capabilities, along with high efficiency, high dynamic range and low noise characteristics are utilised in a wide array of scientific applications ranging from astrophysics~\cite{romani_first_1999, holland_scuba-2_2013, gottardi_review_2021}, particle physics~\cite{irwin_transition-edge_2005, ullom_review_2015} and quantum physics. In the field of quantum information science in particular, the availability of PNR detectors has enabled advances in topics such as quantum metrology and sensing~\cite{xiang2011entanglement, datta_quantum_2011, thekkadath_quantum-enhanced_2020, GerritsMetrology2021}, bio-imaging\cite{fukuda2021},  quantum computing~\cite{Aaronson_Arkhipov_2011, hamilton_gaussian_2017, kruse_detailed_2019, madsen_quantum_2022}, quantum communication~\cite{Smith_conclusive_steering_2012,GiustinaBellTest,Mycroft_proposal-2023}, and quantum state preparation~\cite{gerrits-2010-generation-cat-state, Namekata2010_non_gaussian_operation, Thekkadath2020engineering, magana2019multiphoton, tzitrin_progress_2020, Endo2023_non_gaussian_generation}. 
A key drawback to the TES technology is its slower operation rate compared to other prevalent detectors such as superconducting nanowire single-photon detectors (SNSPDs). The temperature response of a TES to a photon detection event has both a relatively long rise time, primarily limited by the inductance of the electric read-out circuit, and subsequently, a long recovery time, which is limited by the weak electron-phonon coupling between the active device material and the substrate. While the TES is still able to produce an electronic response to subsequent photon absorption events during its post-detection recovery times, this takes place at an elevated base temperature due to the un-dissipated heat. The analogue voltage signals, for the same photon number, would have different shapes depending on the influence of the preceding signals, thereby creating a much more difficult classification problem in signal processing. The presence of background optical and electronic noise further corrupts the signal shape and complicates the classification problem.  

Consequently, methods that classify TES signals by a single attribute, such as pulse height or area~\cite{levine_algorithm_2012, madsen_quantum_2022, morais_precisely_2022}, have poor accuracy if the detector is operated above its thermal recovery rate. For example, in a recent experiment that deployed TESs as PNR detectors for a quantum computational advantage demonstration, the detection rate was limited to 375~kHz despite the photon source operating at a 6~MHz repetition rate~\cite{madsen_quantum_2022}.
Active demultiplexing had to be deployed to reduce the clock rate, which adds additional experimental complexity and hardware overheads, typically at the cost of photon throughput. For larger scale experiments, it is important to mitigate these effects to reach quantum advantage.

Recent fabrication improvements have reduced the TES thermal recovery time by increasing the thermal conductance between the sensor and the substrate using a layer of gold. This pushes the TES operation rate to 1-1.5~MHz, but further speed-up with this approach is traded off against decreasing photon-number resolution~\cite{hummatov_fast_2023, pepe_development_2024}. 
Alternatively, further improvement to the detection rate can be attained by better algorithmic photon-number assignment to the noisy and overlapping TES signals at high repetition rates. 
In this work, we utilise the tool of machine learning to solve this classification problem. We first propose a supervised learning method where a classifier is trained by low-repetition-rate traces and applied to predict the labels of high-repetition-rate signals. We also propose an unsupervised method using a hierarchical density-based clustering algorithm to classify the high-repetition-rate signals directly. We explain how Principal Component Analysis (PCA) can be utilised to reduce the dimensionality of the data. By benchmarking the methods with the detection of coherent states and squeezed states, we show that appropriate selection of algorithm along with its hyperparameters enables an 800~kHz detection rate with accurate photon-number assignment to up to at least five photons, which is quadruple the detectors' intrinsic thermal recovery rate and beats the state-of-the-art result in Ref.~\cite{madsen_quantum_2022}. Our methods provide a hardware-agnostic approach to increasing the bandwidth of technologies that rely on TESs for accurate PNR detection. 


\begin{figure}[!t]
    \centering
    \includegraphics[width=\textwidth]{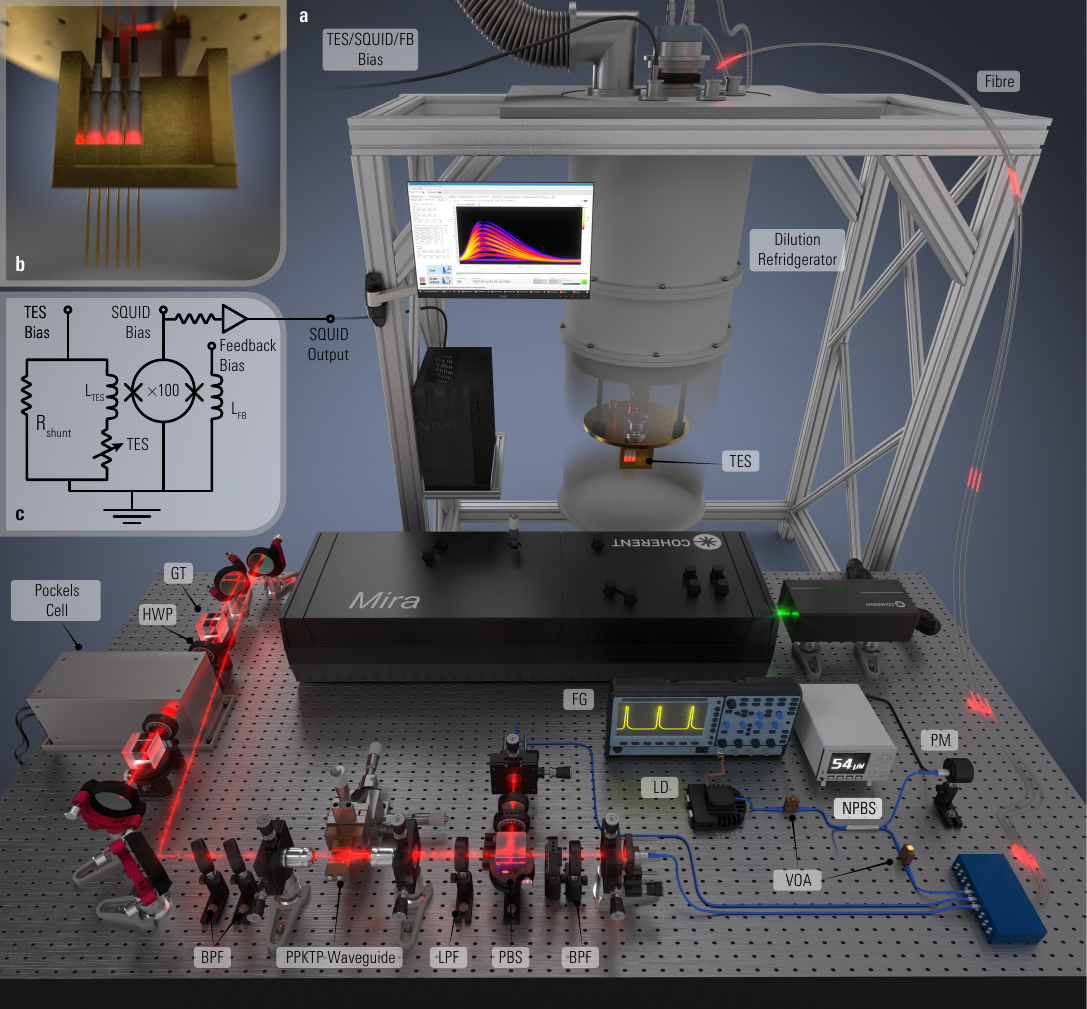}
    \caption{(a) The coherent and squeezed light sources and the PNRD apparatus. Coherent states at a desired repetition rate are generated by modulating a 1550~nm laser diode (LD) with a function generator (FG). Variable optical attenuators (VOA) and a 99:1 non-polarising beamsplitter (NPBS) split the output of the diode into two where one output is attenuated to the few-photon level and the other is measured on a power meter (PM). For squeezed light generation, a Ti:Sapphire laser produces 150~fs long pulses at 80~MHz which are double-passed through a pulse-picker consisting of a Pockels cell sandwiched between half-wave plates (HWP) and orthogonal Glan-Taylor polarisers (GT). The Pockels cell is driven by amplified RF signals from a function generator (not shown). The pump pulses are filtered to 780$\pm$1~nm (full width at half maximum) using a pair of angle-tuned bandpass filters (BPF). The pulses are coupled into a 10~mm long ppKTP waveguide that is phase-matched for type-II PDC. The residual pump is filtered using a long-pass filter (LPF) and a polarising beamsplitter (PBS) separates signal and idler pulses along separate paths, each containing a HWP and a BPF with a matched bandwidth to eliminate spectral correlations between signal and idler. Photons from either source are coupled into the transition-edge sensors (TES), via single-mode fibre, housed in a dilution refrigerator at 130~mK shown in (b). (c) Circuit diagram showing how the TES is coupled to the readout electronics consisting of an array of one hundred SQUIDs. See Methods for more details.
    }
    \label{fig: setup}
\end{figure}

\section{PNR detection by TES}

We first offer a brief overview of our experimental arrangement for generating and detecting multi-photon states of light. Figure \ref{fig: setup}(a) shows the overall arrangement which includes an attenuated laser diode for producing coherent states, and a mode-locked Ti:Sapphire laser system and pulse picker for producing two-mode squeezed vacuum (TMSV) states via the parametric downconversion (PDC) process. Both can be produced at different repetition rates. The generated light, which is in the telecom C-band, is coupled into single-mode fibre and fed into a dilution refrigerator with a base temperature $<130\textrm{ mK}$ where the TESs are mounted, as shown in Fig. \ref{fig: setup}(b). 

The active device material in the TES is a superconducting thin film, such as tungsten for the detectors utilised in this work. The material is temperature-biased near its superconducting critical temperature. A temperature rise, even as small as one caused by the absorption of a single photon, generates a steep and proportionate rise in the material's electrical resistance. This response is amplified by Superconducting QUantum Interference Devices (SQUIDs, Fig. \ref{fig: setup}(c)) that also operate at cryogenic temperatures, and further amplified by low-noise room-temperature electronics. See Methods for further details regarding the experimental setup and the TESs.

\begin{figure}[t]
    \centering
    \includegraphics[width=1\textwidth]{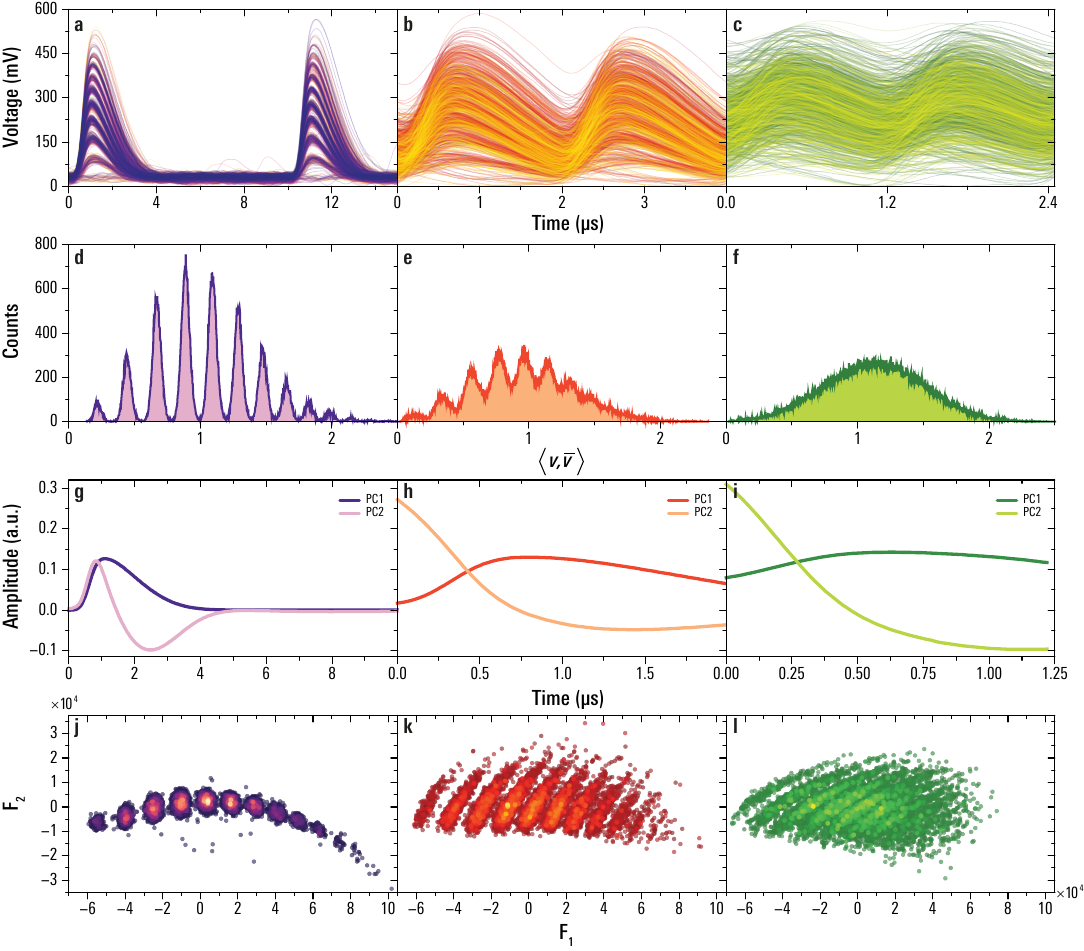}
    \caption{TES and PCA data for coherent states measured at rates 100~kHz (left panels), 500~kHz (middle panels), and 800~kHz (right panels), generated by the light source in Fig. \ref{fig: setup} with equal average power per pulse. (a-c) Raw SQUID signal traces after amplification by low-noise room temperature electronics. The traces are stacked to visualise their clustering into photon-number groups. 
    (d-f) The inner-product histograms that can be used to classify the voltage traces.
    (g-i) The leading two principal components for the mean-subtracted voltage traces. 
    (j-l) A density scatter plot of the voltage traces, where coordinates are given by their first two factor scores from PCA. A lighter colour represents a higher density of points. 
    }
    \label{fig: traces and pca}
\end{figure}

The amplified output from the SQUIDs is digitised into a time-series of voltage values, which can be cut into segments that match the pulse repetition rate. Each segment, which is a vector of voltage values, defines a voltage trace corresponding to a single pulse detection event with examples given in Fig. \ref{fig: traces and pca}(a-c). If adequate time is allowed between two consecutive light pulses, the sensor's temperature and electrical resistance will recover to its baseline value before absorbing the next pulse. As a result, the voltage traces are well-separated in time and show a clear distinction between different bands of traces (Fig. \ref{fig: traces and pca}(a)). For our detectors, this requires the light pulses to have a repetition rate of no more than 200~kHz. 

\subsection{Pulse Filtering}

In a final step known as `pulse filtering', a photon number is assigned to each light pulse can be retrieved by various signal processing methods. One can perform this assignment by an `inner product' (IP) method, which filters the voltage traces by their overlap with a reference trace and shows high photon-number resolution~\cite{levine_algorithm_2012}. Each voltage trace, as a vector of voltage values, is denoted $\bm{v}_i$. We take the average trace to be a reference, $\bm{\bar v}$, and calculate the inner product between each trace with this reference trace, $\left\langle {\bm{v}_i ,\bm{\bar v} } \right\rangle$. The histogram of these inner product values shows distinct peaks with high extinction ratio (Fig. \ref{fig: traces and pca}(d)). The traces can then be appropriately assigned a photon-number label according to the grouping of their inner products. 

However, for detection of higher-repetition-rate light pulses, the voltage traces start to overlap in time. The extinction ratio of their inner-product histograms drastically reduces until no clear peaks can be seen, thereby rendering the method inaccurate or unusable (Fig. \ref{fig: traces and pca}(b-c, e-f)). Similar signal processing methods based on filtering the voltage traces by a single attribute, such as trace height or area under curve~\cite{morais_precisely_2022, madsen_quantum_2022}, would encounter the same issue as the traces become increasingly similar at higher repetition rate. 

\subsection{Principal Component Analysis} 

Another way to visualise the clustering of TES voltage traces is by principal component analysis (PCA)~\cite{abdi_principal_2010, humphreys_tomography_2015}. The voltage traces are decomposed into a weighted sum of `principal components', where each component extracts the largest variance from the data while being orthogonal to the previous components. 

To extract the principal components from a set of signals, we express the full set of signals as a matrix, $\bm{V}$, where each row vector is a voltage trace, $\bm{v}_i$. The average trace,  $\bm{\bar v}$, is subtracted from $\bm{V}$, and the mean-subtracted voltage traces form a new matrix, $\tilde{\bm{V}}$, that has singular value decomposition, $\tilde{\bm{V}} = \bm{P \Delta Q}^T$, where $\bm{P}$ and $\bm{Q}$ are the matrices of left and right singular vectors and $\bm{\Delta}$ the diagonal matrix of singular values. For PCA, matrix $\bm{Q}$ becomes the projection matrix, whose column vectors, $\bm{q}_j$, are the principal components. Each voltage trace can now be expressed as a weighted sum of principal components,
\begin{equation}\label{eqn: pca result on voltage trace}
    \bm{v}_i = \sum_{j=1}^D F_{ij} \bm{q}_j + \bm{\bar v},
\end{equation}
where the weights, $F_{ij}$, known as `factor scores', are elements of the matrix $\bm{F}=\bm{P\Delta}$. The summation upper limit, $D$, is the dimension of the signal, i.e. the length of $\bm{v}_i$~\cite{abdi_principal_2010}. 

Due to the nature of PCA, the summation in Equation \ref{eqn: pca result on voltage trace} can be truncated to a low order while maintaining a good approximation to the raw signal~\cite{humphreys_tomography_2015}. This offers a way to reduce the dimensionality of the data from full voltage traces to points in $N$-dimensional space, where $N$ is the number of principal components one chooses to preserve. 

Figure \ref{fig: traces and pca}(g-l) show an example with $D=2$, which is already sufficient to distinguish photon-number clusters. At low repetition rate, the first-order component is just the rise and fall of the mean-subtracted voltage response to the current light pulse, and the first-order factor scores alone are enough to classify the traces into distinct clusters. Clusters with smaller first-order factor score values correspond to smaller photon numbers, as these are traces with smaller height. At higher repetition rate, however, the second-order factor scores, representing the weight of the tail of the preceding traces, become more important, though their absolute values are still an order of magnitude smaller than the first-order factor scores. The data still organise into clusters, albeit in a 2-dimensional space that includes the second-order component and with much reduced visibility.

\begin{figure}[!t]
    \centering
    \includegraphics{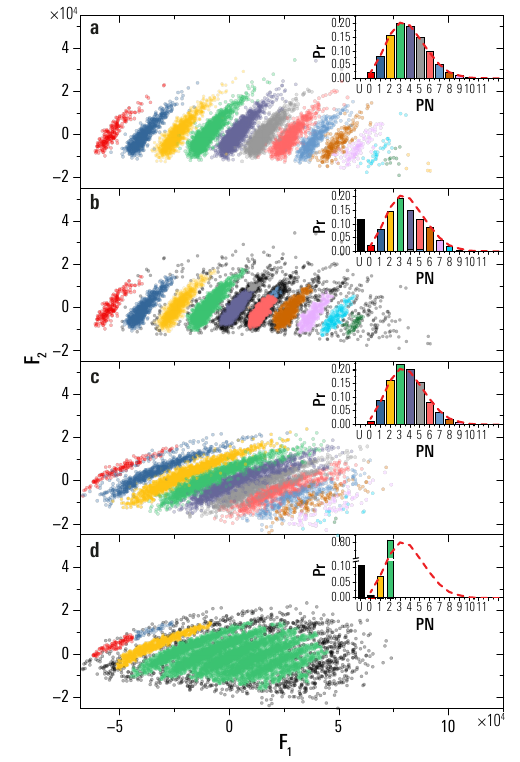}
    \caption{Classification of coherent light pulses at 500~kHz (which are shown in the middle panels of Fig. \ref{fig: traces and pca}) by (a) the KNN algorithm and (b) the HDBSCAN algorithm. A subset of the traces are density-scatter plotted according to their PCA decomposition, and each label assignment is represented by a different colour. In KNN, each label corresponds to a unique photon number (PN). In HDBSCAN, sub-clusters might be identified within a photon-number cluster (e.g. purple and grey for PN=4, and pink and light-blue for PN=5). Unclassified traces are assigned the label ``U" and plotted in black. The inset figures plot the predicted probability (Pr) of a particular photon-number outcome. The classification of coherent light pulses at 800~kHz (right panels of Fig. \ref{fig: traces and pca}) by (c) the KNN algorithm and (d) the HDBSCAN algorithm are also shown. Red dashed curves in inset figures show the reference PN distribution, classified by the IP method for 100~kHz coherent light pulses with equal average power per pulse (left panels of Fig. \ref{fig: traces and pca}).
    }
    \label{fig: ml and pca}
\end{figure}

\section{Pulse filtering by Machine Learning}

\subsection{Supervised Learning Method}
\label{section: supervised learning method}

For detection of higher-repetition-rate light pulses, we adopt a supervised learning method for pulse filtering. In this method, we first calibrate the detector with well-separated pulses at 100~kHz, whose photon-number labels can be accurately assigned by the IP method. Next, their voltage traces are overlapped to emulate the detector's response to light pulses at the target repetition rate, producing a labelled training dataset for a machine-learning classifier. The trained classifier is then applied to subsequent measurements at the target repetition rate. 

We will use the K-Nearest Neighbour (KNN) classifier from the Scikit-Learn Python package as an example~\cite{Cover-Hart-nearest-neighbor-1967, scikit-learn}. The classifier stores the instances of the training data, and unknown data are classified by a simple majority vote of the $K$ nearest neighbours. We choose $K$ to be five and select nearest neighbours by the Euclidean distance between two vectors. 

Figure \ref{fig: ml and pca}(a) and (c) show an example of detecting coherent state light pulses at 500~kHz and 800~kHz, respectively. The KNN classifier successfully classifies the light pulses into photon-number clusters up to 13 photons for 500~kHz and 12 photons for 800~kHz. The assignment error, which can be visualised by the points that cross over into a neighbouring cluster with a different colour, increases at higher rate but remains low. Furthermore, the overall photon-number distribution is close to the reference distribution obtained by running the IP method on light pulses at 100~kHz with equal average power per pulse. 

We also tested other supervised classifiers, such as Random Forest (RF)~\cite{Breiman2001Oct-random-forest, scikit-learn}, Support Vector Machine (SVM)~\cite{liblinear-paper, libsvm-paper, scikit-learn}, Gradient-Boosted Decision Tree (GBDT)~\cite{Chen_2016_xgboost} and a Convolutional Neural Network (CNN) model for time-series classification that consists of two convolutional layers and one global average pooling operation~\cite{tensorflow2015-whitepaper, lin2014networknetwork, wang2016timeseriesclassificationscratch, IsmailFawaz2019Jul-deep-learning-for-time-series-classification}. With the same training procedure, we found that the KNN classifier has the least processing time with comparable accuracy. For coherent light pulses at 800~kHz with 3.82 photons per pulse on average, it took 0.79~s to train on 100,000 training traces and 6.8~s to predict 100,000 unknown traces, averaging at 68~$\mu$s per trace. The training and prediction are performed on entire voltage traces. Further speedup can be obtained by reducing the dimensionality of the data by PCA and by parallelisation on batched traces.  

The primary limitation of the supervised method is its reliance on high-quality training data. This stipulates low noise on the 100~kHz calibration data and easy switching of the repetition rate from low to high, ideally while maintaining the same mean photon number per pulse.

\subsection{Unsupervised Clustering Method}
\label{section: unsupervised clustering method}

To remove the requirement for training, we also propose an unsupervised method based on a spatial clustering algorithm, Hierarchical Density-Based Spatial Clustering of Applications with Noise (HDBSCAN)~\cite{Campello-2013-HDBSCAN, mcinnes2017hdbscan}. We reduce the voltage traces to points in a two-dimensional plane by performing PCA and preserving only the first two factor scores. The HDBSCAN algorithm is applied to identify clusters which are areas of high density. To reduce processing time, the clustering model is first trained on a subset of data, then held fixed and applied to predict the cluster label for the rest of the data. It took 1.6~s to train the clustering model on 20,000 traces from squeezed-light and subsequently, on average, 22~$\mu$s to predict an unknown trace. 

Figure \ref{fig: ml and pca}(b) shows the clustering results of HDBSCAN on the same coherent light pulses in Section \ref{section: supervised learning method} at 500~kHz. The main advantage of HDBSCAN is a lower assignment error, as there is little to no cross-overs between different clusters. The trade-off is that many data points in low-density areas, sandwiched by two high-density clusters, cannot be classified into either cluster and are labelled as noise instead.  This skews the overall photon number distribution. Occasionally, the algorithm identifies extraneous sub-clusters inside a photon-number cluster. These sub-clusters can be manually grouped together in an extra post-processing step.

At 800~kHz, the visibility between different photon-number clusters in the PCA plane is greatly reduced (Fig. \ref{fig: ml and pca}(d)). As a result, the algorithm fails to identify any photon number clusters from two photons upwards despite careful tuning of the model's hyperparameters including the minimum cluster size, the minimum number of neighbours to a core point, and the minimum distance between distinct clusters.

\begin{figure}[!t]
    \centering
    \includegraphics{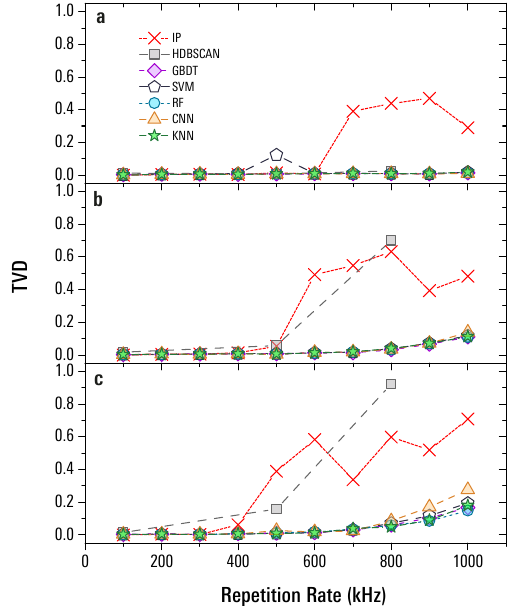}
    \caption{Detection of coherent light with varying mean photon number per pulse, (a) $\mu=0.86$, (b) $\mu=3.82$, (a) $\mu=5.29$. In each case, the TVD is calculated against the reference distribution measured by the inner-product (IP) method at 100~kHz.}
    \label{fig: tvd}
\end{figure}

\begin{figure}[!t]
    \centering
    \includegraphics{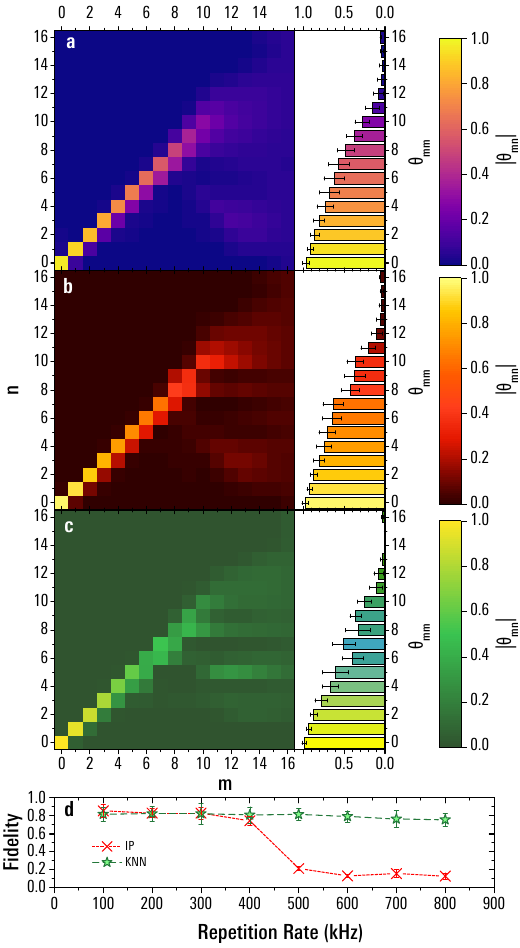}
    \caption{The reconstructed confusion matrix, $\bm{\Theta}$ by the IP and KNN method at (a) 100~kHz, (b) 500~kHz, (c) 800~KHz. Bar plots of the diagonal elements are also plotted. The error bars are estimated by bootstrapping over the uncertainties of the power meter readings and Poissonian photon counting errors. 
    (d) The fidelity of the $\bm{\Theta}$ reconstructions, as defined in Eq. \ref{eqn: define average fidelity}. The reference POVM, $\bm{\Theta}^{(\text{ref})}$, is the averaged reconstruction result by the IP method at 100~kHz.
    }
    \label{fig: tomo}
\end{figure}
\section{Benchmarking}

\subsection{Detection of Coherent Light}
\subsubsection{Total Variation Distance}

To benchmark the performance of our proposed signal processing methods, we first apply them to the detection of coherent state light pulses. The light source is a highly attenuated, modulated, laser diode whose repetition rate can be freely controlled by a function generator. When switching between different rates, we maintain the same pre-attenuation average power per pulse through adjustments of the duty cycle, so as to preserve the photon number distribution. 

Figure \ref{fig: tvd}(a-c) show how the photon number distribution measured by each method deviates from the reference distribution at different repetition rates and pulse power.
The reference distribution is measured by the IP method at 100~kHz repetition rate. 
We use the total variation distance (TVD) as the distance metric between two distributions, defined by $\tvd(\bm{p}, \bm{q})=\sum_i \frac{1}{2}\left|p_i - q_i\right|$. The results confirm the improvement by the supervised machine learning methods over the IP method at higher repetition rates up to at least 800~kHz. The unsupervised clustering method by HDBSCAN, expectedly, has poor accuracy in the photon-number distribution, which is due to the fact that the method classifies a number of traces as `noise' and fails altogether to separate clusters for higher photon numbers at 800~kHz. 

\subsubsection{Detector POVM Reconstruction}
\label{section: detector tomography}
A set of coherent states with varying amplitudes, $\{|\alpha_j\rangle\}$, also forms a tomographically complete set of probe states that allows us to reconstruct the POVM of the phase-insensitive TES, which can be expressed as~\cite{humphreys_tomography_2015, morais_precisely_2022}: 
\begin{equation}\label{eqn: tes povm}
    \{ \hat{\Pi}_{n}: \hat{\Pi}_n = \sum_{m=0}^{+\infty} \theta_{nm} |m\rangle \langle m| \}
\end{equation}
where $|m\rangle$ is an $m$-photon Fock state. In practice, the summation in Eq. \ref{eqn: tes povm} is truncated to some large photon number, $M$. The scalars $\theta_{nm}$, represent the conditional probability that the detector reports $n$-photons given an $m$-photon input and form a confusion matrix $\bm{\Theta}$ that illustrates the accuracy of the detector. 

To reconstruct the detector's POVM in Eq. \ref{eqn: tes povm}, we first calibrate the amplitudes of the probe states, $\{|\alpha_j\rangle\}$. The light from the laser diode is split on a calibrated beamsplitter, with one output being monitored by an independent power meter, and the other output being measured by the TES after a calibrated two-stage fixed attenuation. The amplitude is adjusted by a variable optical attenuator before the beamsplitter. Assuming perfect Poissonian behaviour, the ground-truth distributions of the coherent light can be calculated. One can then reconstruct the confusion matrix $\bm{\Theta}$ by a least-square minimisation procedure between the TES-measured distributions and the ground-truth distributions multiplied by a parameterised version of Eq. \ref{eqn: tes povm}~\cite{humphreys_tomography_2015, morais_precisely_2022}. The optimisation routine was performed using the CVXPY Python package~\cite{diamond2016cvxpy, agrawal2018rewriting}.

Assuming a perfect power meter, the mean photon number detected by the IP method at 100~kHz gives a detection efficiency of 93.3\% at the 1550~nm wavelength (see Supplemental Material). Realistically, however, this efficiency estimate will also include inaccuracy in the power meter reading and calibration errors. 
To benchmark the accuracy of the signal processing methods with minimal error carried over from the power meter's and the detector's intrinsic behaviour, we use the reconstructed POVM by the IP method at 100~kHz as a reference, and compare it with the reconstructed POVMs at higher repetition rates. 

We define the average fidelity between a reconstructed $\bm{\Theta}$ and the reference $\bm{\Theta}^{\text{ref}}$ by~\cite{zhang_mapping_2012}
\begin{equation}\label{eqn: define average fidelity}
    F = \frac{1}{N} \sum_n^N
    \frac{
    \left(
    \sum_m^M \sqrt{\theta_{nm} \theta_{nm}^{(\text{ref})}}
    \right)^2
    }{
    \left(
    \sum_m^M \theta_{nm}
    \right)
    \left(
    \sum_m^M \theta_{nm}^{(\text{ref})}
    \right)
    },
\end{equation}
where the reported and actual photon number are truncated to $N$ and $M$ respectively. The average fidelity with $N=M=16$ shows that the KNN algorithm's performance at up to 800~kHz remains on par with the detector's performance at low repetition rates (Fig. \ref{fig: tomo}(d)). 

Looking more closely at the reconstructed diagonal terms of $\bm{\Theta}$, Fig. \ref{fig: tomo}(a-c, right panels) show that the KNN algorithm preserves the detection accuracy for up to at least seven photons at 500~kHz and five photons at 800~kHz.
A caveat to this is that this detection accuracy will also depend on the state of the input light. For example, while the reconstructed POVM at 500~kHz showed a drop in accuracy for 8-photon detection (Fig. \ref{fig: tomo}(b)), this is an averaged result across the probe states. For probe states with lower mean photon number, the ``tail'' influence from preceding traces is on average smaller, reducing the difficulty of the classification problem. For example, for the light pulses shown in Fig. \ref{fig: ml and pca}(a) which average at 3.82 photons per pulse, the KNN algorithm was still able to label traces accurately up to 13 photons at 500~kHz. 

\subsection{Detection of Squeezed Light}
\label{section: squeezed light detection}

\begin{figure}[!t]
    \centering
    \includegraphics{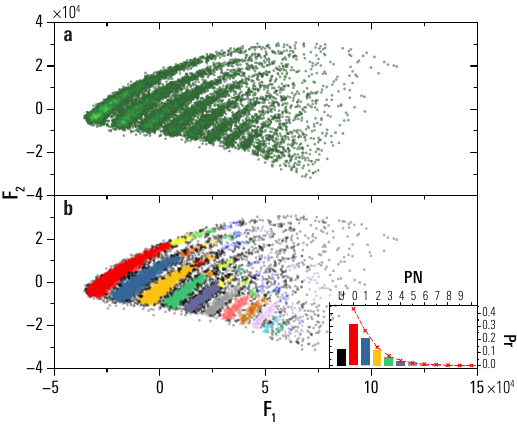}
    \caption{
    (a) Density scatter plot of the PCA factor scores for light pulses in a thermal state. 
    (b) The classification result by the HDBSCAN algorithm visualised on the PCA scatter plot. Each cluster is plotted with a different colour. 
    (inset) The photon-number distribution of the thermal state, where the colour of each bar is the colour of the corresponding cluster in (b). Sub-clusters that have the same photon-number label are grouped together. The red-dashed curve is the distribution reconstructed by the inner product method at 100~kHz for the same average pump power per pulse.}
    \label{fig: Thermal}
\end{figure}

\begin{figure}[!t]
    \centering
    \includegraphics{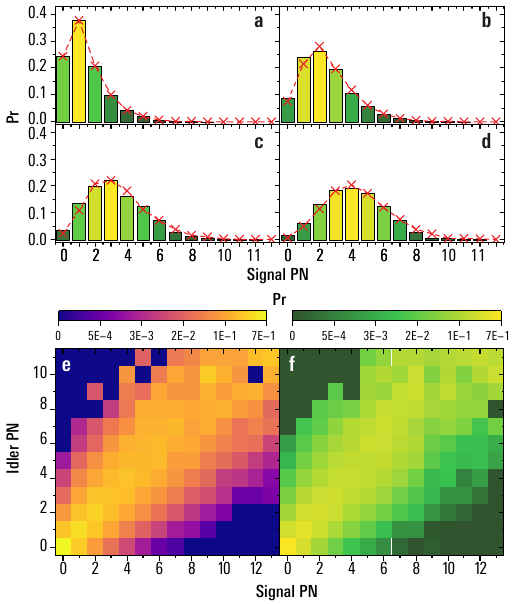}
    \caption{
    The photon-number distribution of the signal mode whilst heralding on (a) one, (b) two, (c) three, and (d) four photons in the idler mode. The red dashed curve and crosses represents the reference distribution obtained by the IP method at 100~kHz for the same average pump power per pulse. 
    The two-mode photon-number distribution reconstructed by (e) the inner product method at 100~kHz and (f) the HDBSCAN algorithm at 800~kHz. All data have been renormalised after discarding the unclassified traces.}
    \label{fig: Fock}
\end{figure}

Squeezed states are reduced quantum uncertainty states that have proved to be an invaluable resource in quantum information and sensing. The ability of performing PNR detection on squeezed light has enabled key applications in topics such as photonic quantum computing~\cite{madsen_quantum_2022}, sub-shot-noise quantum sensing~\cite{thekkadath_quantum-enhanced_2020} and highly non-classical state generation~\cite{gerrits-2010-generation-cat-state, Endo2023_non_gaussian_generation}. With a distinctly different photon-number distribution from the coherent light, detection of squeezed light may pose a different classification task for our algorithms, which we benchmark in this section. 

Our squeezed-light source is based on a periodically-poled potassium titanyl phosphate (ppKTP) waveguide pumped by a pulsed laser at 780$\pm$1~nm wavelength. The waveguide produces orthogonally-polarized signal and idler photons at 1560 nm via the type-II PDC process. The generated pairs exhibit photon-number correlation, forming a two-mode squeezed vacuum (TMSV) state described by: $|\tmsv\rangle =\sqrt{1 - |\lambda|^2} \sum_{n=0}^{\infty} \lambda^n | nn \rangle$, where $\lambda$ is the parameter that determines the mean photon number per pulse and depends on the pump field and the ppKTP crystal. The repetition rate of the pump light is 80~MHz but can be modulated by a Pockels cell, which acts as a voltage-controlled pulse picker. We first collect the 800~kHz data. Then we reduce the repetition rate of the pump light to 100~kHz while maintaining the same average pump power per pulse, and measure the photon number distribution by the IP method, which is used as the reference distribution. 

The signal and idler pulses are split into two spatial modes and coupled into the TESs. In each mode, the TES records a thermal state distribution. As a result, the PCA result has a different density distribution from that for a coherent state (Fig. \ref{fig: Thermal}(a)). The factor scores for the second-order component, which represents the weight of the tail of the preceding traces, are concentrated in lower values, since the thermal distribution is dominated by lower photon numbers. This reduces the difficulty for our unsupervised clustering method. Even at 800~kHz repetition rate, the HDBSCAN algorithm was able to identify clusters corresponding to photon numbers up to 10 (Fig. \ref{fig: Thermal}(b)), which is a drastic improvement from the detection of coherent states.

The algorithm still leaves more than 10\% of the traces unclassified, resulting in an inaccurate reconstruction of the thermal distribution. However, among the traces that the algorithm was able to classify, the assignment error is low, which is illustrated by the low cross-overs between the different clusters in Fig. \ref{fig: Thermal}(b). 

The low assignment error manifests in the preparation of heralded Fock states. The photon-number correlation in a TMSV state allows the preparation of an $n$-photon Fock state in the signal mode by heralding on the idler mode. For up to $n=4$, the heralded Fock state detected by the HDBSCAN algorithm produces the correct photon number with the highest probability. The spread of probabilities in unwanted photon numbers is due to photon loss between the PDC source and the detector, rather than due to the detection scheme. After discarding the unclassified traces and renormalising the remaining probabilities, the heralded photon number distributions at 800~kHz by the HDBSCAN algorithm are almost identical to the reference distributions at 100~kHz (Fig. \ref{fig: Fock}(a-d)). The overall fidelity between the two-mode photon number distribution at 800~kHz and at 100~kHz (Fig. \ref{fig: Fock}(e-f)) is $F=\left(\sum_i \sqrt{p_i q_i}\right)^2=99.2\%$. 

The supervised KNN classification algorithm was also tested on the detection of the TMSV state and reported 97.6\% fidelity at 800~kHz, which is worse than the unsupervised HDBSCAN algorithm. This is likely due to a lower signal-to-noise ratio by the Pockels cell, as compared to the laser diode for coherent light detection, which reduces the quality of the training data and reduces the algorithm's ability to differentiate low-photon-number traces from background noise. 

\section{Conclusions}
In this work, we reported two machine-learning-based signal processing methods for TESs that can extend their PNR capabilities beyond the limitations of their sensors' thermal recovery time: One supervised learning method and one unsupervised clustering method. 
Testing with coherent states, the first method enabled our TES to operate at up to 800~kHz with little-to-no drop in accuracy for the detection of up to five photons, beating the state-of-the-art signal processing method in Ref.~\cite{madsen_quantum_2022}. This is achieved by constructing a training data set from 100~kHz voltage traces, where photon numbers can be accurately assigned via the IP method, and overlapping the traces to emulate a particular detection rate. This is verified against the tomographic reconstruction of the TES's POVM at 100~kHz where the TES voltage traces are well-separated in time.

The unsupervised method removes the requirement for training and clusters the raw voltage traces according to their PCA factor scores. The weakness of this density-based clustering method is traces that do not obviously belong to any cluster might be mislabelled as noise. However, in the traces that the algorithm is able to label, the method has very high precision and low assignment error. This is showcased in the algorithm's ability to prepare heralded Fock states of up to four photons at 800~kHz from a TMSV state of light. 

Our proposed methods provide the algorithmic approach to speed up PNR detection with a TES. The hardware-agnostic nature means that they can be implemented in conjunction with any existing or future fabrication improvements, and deliver the last-mile for speeding up quantum experiments that require PNR capability. Notable examples include applications where operations are conditioned on PNR measurements, requiring fast feedforward of PNR outcomes, such as non-Gaussian quantum state preparation~\cite{gerrits-2010-generation-cat-state, Namekata2010_non_gaussian_operation, Thekkadath2020engineering, magana2019multiphoton, tzitrin_progress_2020, Endo2023_non_gaussian_generation}.
In other fields of research that utilise the TES, such as bio-imaging~\cite{fukuda2021}, higher detection rates may benefit dynamic studies like live-cell imaging or real-time biological processes where fast image acquisition with low signal-to-noise are essential. Overcoming the slow recovery time might also be beneficial for the detection of burst of particles and in studies of the cosmic microwave background~\cite{dutcher2024simons} where fine-temporal resolution of events is desired. 

Finally, we note that the machine-learning algorithms reported here may be further improved by fine-tuning the algorithm's parameters and behaviour to the specific experimental conditions. 

The code developed in this paper is publicly accessible in ref \cite{li_github} and we hope our paper can provide a gateway to further algorithmic improvements on TES signal processing. 

\textit{Note added--We recently became aware of an unsupervised machine learning algorithm~\cite{dalbec2024accurate}, for the labelling of signals, which may also be applicable to temporally overlapping signals.
}
\setcounter{section}{0}
\renewcommand{\thesection}{Appendix \Alph{section}}
\renewcommand{\thesubsection}{\arabic{subsection}}
\section{Methods}
\subsection{Experimental set-up}

The experimental setup is illustrated in Fig. \ref{fig: setup}(a). The light source for generating coherent states is based on a 1550~nm wavelength pigtailed fibre laser diode driven by a function generator producing an RF signal consisting of 70~ns wide square pulses. The output light is reduced to the few-photon level by three stages of fibre-based attenuation: A variable optical attenuator, a non-polarising beamsplitter with 99:1 splitting ratio, and two optical attenuators (only one shown) connected in series with 30~dB attenuation each. The beamsplitter output with the larger intensity is monitored by a Newport 918-IG-OD3R power meter. The amplitudes of the coherent states are varied by the first variable optical attenuator. The repetition rate of the signal is controlled by the function generator. When changing between different repetition rates, the amplitude of the RF signal is adjusted such that the power meter records the same average power per pulse to compensate for the differences in rise and fall times of the square pulses. 

For the tomography routine, the amplitudes of the probe states $\{|\alpha_j\rangle\}$ need to be precisely known. Calibration of the asymmetric beamsplitter and the attenuation scheme is performed by the same Newport 918-IG-OD3R power meter, which specifies 2.6\% reading uncertainty and 0.5\% linearity. However, since the power meter has not been recalibrated since purchase, the calibration uncertainty could be higher. 

For Section \ref{section: squeezed light detection}, a Parametric Down-Conversion (PDC) source was used to generate the two-mode squeezed vacuum (TMSV) state. The PDC source consists of a periodically-poled KTP (ppKTP) waveguide pumped by a femtosecond mode-locked Ti:Sapphire laser (Coherent Mira-HP, pumped by a Verdi G18). The pump pulses are filtered to 780 $\pm$ 1~nm (full-width at half-maximum) using a pair of angle-tuned bandpass filters. The Type-II PDC process generates orthogonally-polarized signal and idler modes in the same spatial mode, which are spatially separated by a polarising beamsplitter (PBS) and directed through single-mode optical fibres into two TES detectors. To enhance the spectral purity of the TMSV state, a bandpass filter is placed in front of each fiber coupler. 

The inherent repetition rate of the Ti:Sapphire pump laser is 80~MHz. Pulse-picking is performed by double-passing the pump light through a Pockels cell containing an RTP crystal. The Pockels cell acts as a fast, switchable, HWP rotating the polarisation of the incident pump pulse by $\frac{\pi}{2}$. Sandwiched between orthogonal Glan-Taylor (GT) polarisers and HWPs, the Pockels cell reduces the effective repetition rate of the Ti:Sapphire laser system to $\leq 1\textrm{ MHz}$, with 40~dB extinction. 

\subsection{TES detectors}
The active material in the TES is a tungsten thin film sandwiched between a multi-layer optical structure to optimise absorption efficiency at 1550~nm wavelength. For this type of structure, above 95\% detection efficiency was reported by NIST~\cite{lita_counting_2008}. Our measurement showed a 93.3\% efficiency (see Supplemental Material), though this figure was calculated against an uncalibrated power meter (Newport 918-IG-OD3R). As shown in Fig.~\ref{fig: setup}(a) and (b), the TESs are housed inside an Oxford Instruments Triton 200 dilution refrigerator whose base temperature is maintained below 130~mK.

Each individual TES is connected in series with an inductor and in parallel with a shunt resistor (Fig.~\ref{fig: setup}(c)). The voltage-biased sensors are maintained at the material's superconducting transition phase by the negative electro-thermal feedback effect~\cite{irwin_application_1995}. Upon absorbing photons, the sensor steeply rises in temperature, as well as resistance. This reduces the current flowing through the inductor, causing a change in magnetic flux that is amplified by a SQUID array. The SQUIDs output a $\mu$V-level voltage signal that is further amplified to mV-level by low-noise amplifiers ($4~\textrm{nV}/\sqrt{\textrm{Hz}}$ RMS) at room temperature. To increase the signal-to-noise ratio, a 1~MHz low-pass filter is used to filter out background noise. Finally, the voltage traces are collected by a 14-bit waveform digitiser (AlazarTech ATS9440) for processing on an AMD Threadripper 7970X-based workstation.

\begin{backmatter}

\bmsection{Acknowledgments}
We thank the National Institute of Standards and Technology (NIST) for providing the TESs and SQUIDs used in this work, and Adriana Lita for useful discussions regarding the TES recovery process.

This work is partly funded by Engineering and Physical Sciences Research Council and Quantum Computing and Simulation Hub (project T001062), UK Research and Innovation Future Leaders Fellowship (project MR/W011794/1), National Research Council of Canada (project QSP 062-2), EU Horizon 2020 Marie Sklodowska-Curie Innovation Training Network (project no. 956071, `AppQInfo'), and UK Research and Innovation Guarantee Postdoctoral Fellowship (project EP/Y029631/1).

\bmsection{Disclosures}
The authors declare no conflicts of interest.






\bmsection{Data availability} Data underlying the results presented in this paper are not publicly available at this time but may be obtained from the authors upon reasonable request.

\bmsection{Supplemental document}
See Supplemental Material for supporting content. 

\end{backmatter}


\bibliography{references}

\begin{thebibliography}{10}
\newcommand{\enquote}[1]{``#1''}

\bibitem{lita_counting_2008}
A.~E. Lita, A.~J. Miller, and S.~W. Nam, \enquote{Counting near-infrared single-photons with 95\% efficiency,} {\protect\JournalTitle{Optics Express}} \textbf{16}, 3032 (2008).

\bibitem{lita_superconducting_2010}
A.~E. Lita, B.~Calkins, L.~A. Pellouchoud, \emph{et~al.}, \enquote{{Superconducting transition-edge sensors optimized for high-efficiency photon-number resolving detectors},} in \emph{Advanced Photon Counting Techniques IV,}  vol. 7681 M.~A. Itzler and J.~C. Campbell, eds., International Society for Optics and Photonics (SPIE, 2010), p. 76810D.

\bibitem{romani_first_1999}
R.~W. Romani, A.~J. Miller, B.~Cabrera, \emph{et~al.}, \enquote{First {Astronomical} {Application} of a {Cryogenic} {Transition} {Edge} {Sensor} {Spectrophotometer},} {\protect\JournalTitle{The Astrophysical Journal}} \textbf{521}, L153--L156 (1999).

\bibitem{holland_scuba-2_2013}
W.~S. Holland, D.~Bintley, E.~L. Chapin, \emph{et~al.}, \enquote{{SCUBA}-2: the 10 000 pixel bolometer camera on the {James} {Clerk} {Maxwell} {Telescope},} {\protect\JournalTitle{Monthly Notices of the Royal Astronomical Society}} \textbf{430}, 2513--2533 (2013).

\bibitem{gottardi_review_2021}
L.~Gottardi and K.~Nagayashi, \enquote{A {Review} of {X}-ray {Microcalorimeters} {Based} on {Superconducting} {Transition} {Edge} {Sensors} for {Astrophysics} and {Particle} {Physics},} {\protect\JournalTitle{Applied Sciences}} \textbf{11}, 3793 (2021).

\bibitem{irwin_transition-edge_2005}
K.~Irwin and G.~Hilton, \enquote{Transition-{Edge} {Sensors},} in \emph{Cryogenic {Particle} {Detection},}  C.~Enss, ed. (Springer, Berlin, Heidelberg, 2005), pp. 63--150.

\bibitem{ullom_review_2015}
J.~N. Ullom and D.~A. Bennett, \enquote{Review of superconducting transition-edge sensors for x-ray and gamma-ray spectroscopy,} {\protect\JournalTitle{Superconductor Science and Technology}} \textbf{28}, 084003 (2015).

\bibitem{xiang2011entanglement}
G.-Y. Xiang, B.~L. Higgins, D.~Berry, \emph{et~al.}, \enquote{Entanglement-enhanced measurement of a completely unknown optical phase,} {\protect\JournalTitle{Nature Photonics}} \textbf{5}, 43--47 (2011).

\bibitem{datta_quantum_2011}
A.~Datta, L.~Zhang, N.~Thomas-Peter, \emph{et~al.}, \enquote{Quantum metrology with imperfect states and detectors,} {\protect\JournalTitle{Physical Review A}} \textbf{83}, 063836 (2011).

\bibitem{thekkadath_quantum-enhanced_2020}
G.~S. Thekkadath, M.~E. Mycroft, B.~A. Bell, \emph{et~al.}, \enquote{Quantum-enhanced interferometry with large heralded photon-number states,} {\protect\JournalTitle{npj Quantum Information}} \textbf{6}, 89 (2020).

\bibitem{GerritsMetrology2021}
C.~You, M.~Hong, P.~Bierhorst, \emph{et~al.}, \enquote{Scalable multiphoton quantum metrology with neither pre- nor post-selected measurements,} {\protect\JournalTitle{Applied Physics Reviews}} \textbf{8}, 041406 (2021).

\bibitem{fukuda2021}
K.~Niwa, K.~Hattori, and D.~Fukuda, \enquote{Few-photon spectral confocal microscopy for cell imaging using superconducting transition edge sensor,} {\protect\JournalTitle{Frontiers in Bioengineering and Biotechnology}} \textbf{9} (2021).

\bibitem{Aaronson_Arkhipov_2011}
S.~Aaronson and A.~Arkhipov, \enquote{The computational complexity of linear optics,} in \emph{Proceedings of the forty-third annual ACM symposium on Theory of computing,}  (Association for Computing Machinery, New York, NY, USA, 2011), STOC ’11, p. 333–342.

\bibitem{hamilton_gaussian_2017}
C.~S. Hamilton, R.~Kruse, L.~Sansoni, \emph{et~al.}, \enquote{Gaussian {Boson} {Sampling},} {\protect\JournalTitle{Physical Review Letters}} \textbf{119}, 170501 (2017).

\bibitem{kruse_detailed_2019}
R.~Kruse, C.~S. Hamilton, L.~Sansoni, \emph{et~al.}, \enquote{Detailed study of {Gaussian} boson sampling,} {\protect\JournalTitle{Physical Review A}} \textbf{100}, 032326 (2019).

\bibitem{madsen_quantum_2022}
L.~S. Madsen, F.~Laudenbach, M.~F. Askarani, \emph{et~al.}, \enquote{Quantum computational advantage with a programmable photonic processor,} {\protect\JournalTitle{Nature}} \textbf{606}, 75--81 (2022).

\bibitem{Smith_conclusive_steering_2012}
D.~H. Smith, G.~Gillett, M.~P. de~Almeida, \emph{et~al.}, \enquote{{Conclusive quantum steering with superconducting transition-edge sensors},} {\protect\JournalTitle{Nat. Commun.}} \textbf{3}, 1--6 (2012).

\bibitem{GiustinaBellTest}
M.~Giustina, M.~A.~M. Versteegh, S.~Wengerowsky, \emph{et~al.}, \enquote{Significant-loophole-free test of bell's theorem with entangled photons,} {\protect\JournalTitle{Phys. Rev. Lett.}} \textbf{115}, 250401 (2015).

\bibitem{Mycroft_proposal-2023}
M.~E. Mycroft, T.~McDermott, A.~Buraczewski, and M.~Stobi\ifmmode~\acute{n}\else \'{n}\fi{}ska, \enquote{Proposal for the distribution of multiphoton entanglement with optimal rate-distance scaling,} {\protect\JournalTitle{Phys. Rev. A}} \textbf{107}, 012607 (2023).

\bibitem{gerrits-2010-generation-cat-state}
T.~Gerrits, S.~Glancy, T.~S. Clement, \emph{et~al.}, \enquote{Generation of optical coherent-state superpositions by number-resolved photon subtraction from the squeezed vacuum,} {\protect\JournalTitle{Phys. Rev. A}} \textbf{82}, 031802 (2010).

\bibitem{Namekata2010_non_gaussian_operation}
N.~Namekata, Y.~Takahashi, G.~Fujii, \emph{et~al.}, \enquote{{Non-Gaussian operation based on photon subtraction using a photon-number-resolving detector at a telecommunications wavelength},} {\protect\JournalTitle{Nat. Photonics}} \textbf{4}, 655--660 (2010).

\bibitem{Thekkadath2020engineering}
G.~S. Thekkadath, B.~A. Bell, I.~A. Walmsley, and A.~I. Lvovsky, \enquote{Engineering {S}chr{\"{o}}dinger cat states with a photonic even-parity detector,} {\protect\JournalTitle{{Quantum}}} \textbf{4}, 239 (2020).

\bibitem{magana2019multiphoton}
O.~S. Maga{\~n}a-Loaiza, R.~d.~J. Le{\'o}n-Montiel, A.~Perez-Leija, \emph{et~al.}, \enquote{Multiphoton quantum-state engineering using conditional measurements,} {\protect\JournalTitle{npj Quantum Information}} \textbf{5}, 80 (2019).

\bibitem{tzitrin_progress_2020}
I.~Tzitrin, J.~E. Bourassa, N.~C. Menicucci, and K.~K. Sabapathy, \enquote{Progress towards practical qubit computation using approximate {Gottesman}-{Kitaev}-{Preskill} codes,} {\protect\JournalTitle{Physical Review A}} \textbf{101}, 032315 (2020).

\bibitem{Endo2023_non_gaussian_generation}
M.~Endo, R.~He, T.~Sonoyama, \emph{et~al.}, \enquote{{Non-Gaussian quantum state generation by multi-photon subtraction at the telecommunication wavelength},} {\protect\JournalTitle{Opt. Express}} \textbf{31}, 12865--12879 (2023).

\bibitem{levine_algorithm_2012}
Z.~H. Levine, T.~Gerrits, A.~L. Migdall, \emph{et~al.}, \enquote{Algorithm for finding clusters with a known distribution and its application to photon-number resolution using a superconducting transition-edge sensor,} {\protect\JournalTitle{Journal of the Optical Society of America B}} \textbf{29}, 2066 (2012).

\bibitem{morais_precisely_2022}
L.~A. Morais, T.~Weinhold, M.~P.~d. Almeida, \emph{et~al.}, \enquote{Precisely determining photon-number in real time,} {\protect\JournalTitle{Quantum}} \textbf{8}, 1355 (2024).

\bibitem{hummatov_fast_2023}
R.~Hummatov, A.~E. Lita, T.~Farrahi, \emph{et~al.}, \enquote{Fast transition-edge sensors suitable for photonic quantum computing,} {\protect\JournalTitle{Journal of Applied Physics}} \textbf{133}, 234502 (2023).

\bibitem{pepe_development_2024}
C.~Pepe, \enquote{Development of superconducting single-particle detector {Transition}-{Edge} {Sensor},} Ph.D. thesis, Politecnico di Torino, Torino, Italy (2024).

\bibitem{abdi_principal_2010}
H.~Abdi and L.~J. Williams, \enquote{Principal component analysis,} {\protect\JournalTitle{WIREs Computational Statistics}} \textbf{2}, 433--459 (2010).

\bibitem{humphreys_tomography_2015}
P.~C. Humphreys, B.~J. Metcalf, T.~Gerrits, \emph{et~al.}, \enquote{Tomography of photon-number resolving continuous-output detectors,} {\protect\JournalTitle{New Journal of Physics}} \textbf{17}, 103044 (2015).

\bibitem{Cover-Hart-nearest-neighbor-1967}
T.~Cover and P.~Hart, \enquote{Nearest neighbor pattern classification,} {\protect\JournalTitle{IEEE Transactions on Information Theory}} \textbf{13}, 21--27 (1967).

\bibitem{scikit-learn}
F.~Pedregosa, G.~Varoquaux, A.~Gramfort, \emph{et~al.}, \enquote{Scikit-learn: Machine learning in {P}ython,} {\protect\JournalTitle{Journal of Machine Learning Research}} \textbf{12}, 2825--2830 (2011).

\bibitem{Breiman2001Oct-random-forest}
L.~Breiman, \enquote{{Random Forests},} {\protect\JournalTitle{Machine Learning}} \textbf{45}, 5--32 (2001).

\bibitem{liblinear-paper}
R.-E. Fan, K.-W. Chang, C.-J. Hsieh, \emph{et~al.}, \enquote{Liblinear: A library for large linear classification,} {\protect\JournalTitle{J. Mach. Learn. Res.}} \textbf{9}, 1871–1874 (2008).

\bibitem{libsvm-paper}
C.-C. Chang and C.-J. Lin, \enquote{Libsvm: A library for support vector machines,} {\protect\JournalTitle{ACM Trans. Intell. Syst. Technol.}} \textbf{2} (2011).

\bibitem{Chen_2016_xgboost}
T.~Chen and C.~Guestrin, \enquote{Xgboost: A scalable tree boosting system,} in \emph{Proceedings of the 22nd ACM SIGKDD International Conference on Knowledge Discovery and Data Mining,}  (ACM, 2016), KDD ’16, p. 785–794.

\bibitem{tensorflow2015-whitepaper}
M.~Abadi, A.~Agarwal, P.~Barham, \emph{et~al.}, \enquote{{TensorFlow}: Large-scale machine learning on heterogeneous systems,}  (2015). Software available from tensorflow.org.

\bibitem{lin2014networknetwork}
M.~Lin, Q.~Chen, and S.~Yan, \enquote{{Network In Network},} {\protect\JournalTitle{arXiv preprint arxiv:1312.4400}}  (2013).

\bibitem{wang2016timeseriesclassificationscratch}
Z.~Wang, W.~Yan, and T.~Oates, \enquote{Time series classification from scratch with deep neural networks: A strong baseline,} in \emph{2017 International joint conference on neural networks (IJCNN),}  (IEEE, 2017), pp. 1578--1585.

\bibitem{IsmailFawaz2019Jul-deep-learning-for-time-series-classification}
H.~Ismail~Fawaz, G.~Forestier, J.~Weber, \emph{et~al.}, \enquote{{Deep learning for time series classification: a review},} {\protect\JournalTitle{Data Min. Knowl. Disc.}} \textbf{33}, 917--963 (2019).

\bibitem{Campello-2013-HDBSCAN}
R.~J. G.~B. Campello, D.~Moulavi, and J.~Sander, \enquote{Density-based clustering based on hierarchical density estimates,} in \emph{Advances in Knowledge Discovery and Data Mining,}  J.~Pei, V.~S. Tseng, L.~Cao, \emph{et~al.}, eds. (Springer Berlin Heidelberg, Berlin, Heidelberg, 2013), pp. 160--172.

\bibitem{mcinnes2017hdbscan}
L.~McInnes, J.~Healy, and S.~Astels, \enquote{hdbscan: Hierarchical density based clustering,} {\protect\JournalTitle{The Journal of Open Source Software}} \textbf{2}, 205 (2017).

\bibitem{diamond2016cvxpy}
S.~Diamond and S.~Boyd, \enquote{{CVXPY}: {A} {P}ython-embedded modeling language for convex optimization,} {\protect\JournalTitle{Journal of Machine Learning Research}} \textbf{17}, 1--5 (2016).

\bibitem{agrawal2018rewriting}
A.~Agrawal, R.~Verschueren, S.~Diamond, and S.~Boyd, \enquote{A rewriting system for convex optimization problems,} {\protect\JournalTitle{Journal of Control and Decision}} \textbf{5}, 42--60 (2018).

\bibitem{zhang_mapping_2012}
L.~Zhang, H.~B. Coldenstrodt-Ronge, A.~Datta, \emph{et~al.}, \enquote{Mapping coherence in measurement via full quantum tomography of a hybrid optical detector,} {\protect\JournalTitle{Nature Photonics}} \textbf{6}, 364--368 (2012).

\bibitem{dutcher2024simons}
D.~Dutcher, S.~M. Duff, J.~C. Groh, \emph{et~al.}, \enquote{The simons observatory: large-scale characterization of 90/150 {GH}z {TES} detector modules,} {\protect\JournalTitle{Journal of Low Temperature Physics}} \textbf{214}, 247--255 (2024).

\bibitem{li_github}
Z.~Li and M.~Kendall, \enquote{{High\_speed\_TES\_ML},} \url{https://github.com/apprenticeadi/High_speed_TES_ML} (2024).

\bibitem{dalbec2024accurate}
N.~Dalbec-Constant, G.~Thekkadath, D.~England, \emph{et~al.}, \enquote{Accurate unsupervised photon counting from transition edge sensor signals,} {\protect\JournalTitle{arXiv preprint arXiv:2411.05737}}  (2024).

\bibitem{irwin_application_1995}
K.~D. Irwin, \enquote{An application of electrothermal feedback for high resolution cryogenic particle detection,} {\protect\JournalTitle{Applied Physics Letters}} \textbf{66}, 1998--2000 (1995).

\bibitem{2020SciPy-NMeth}
P.~Virtanen, R.~Gommers, T.~E. Oliphant, \emph{et~al.}, \enquote{{{SciPy} 1.0: Fundamental Algorithms for Scientific Computing in Python},} {\protect\JournalTitle{Nature Methods}} \textbf{17}, 261--272 (2020).

\end{thebibliography}

\pagebreak
\begin{center}
\textbf{\large Supplemental Material: Boosting Photon-Number-Resolved Detection Rates of Transition-Edge Sensors by Machine Learning}
\end{center}
\setcounter{equation}{0}
\setcounter{figure}{0}
\setcounter{section}{0}
\makeatletter

\renewcommand{\theequation}{S\arabic{equation}}
\renewcommand{\thesection}{S\Roman{section}}
\renewcommand{\thefigure}{S\arabic{figure}}

\section{Detector efficiency}

\begin{figure}[!b]
    \centering
    \includegraphics[width=1.\textwidth]{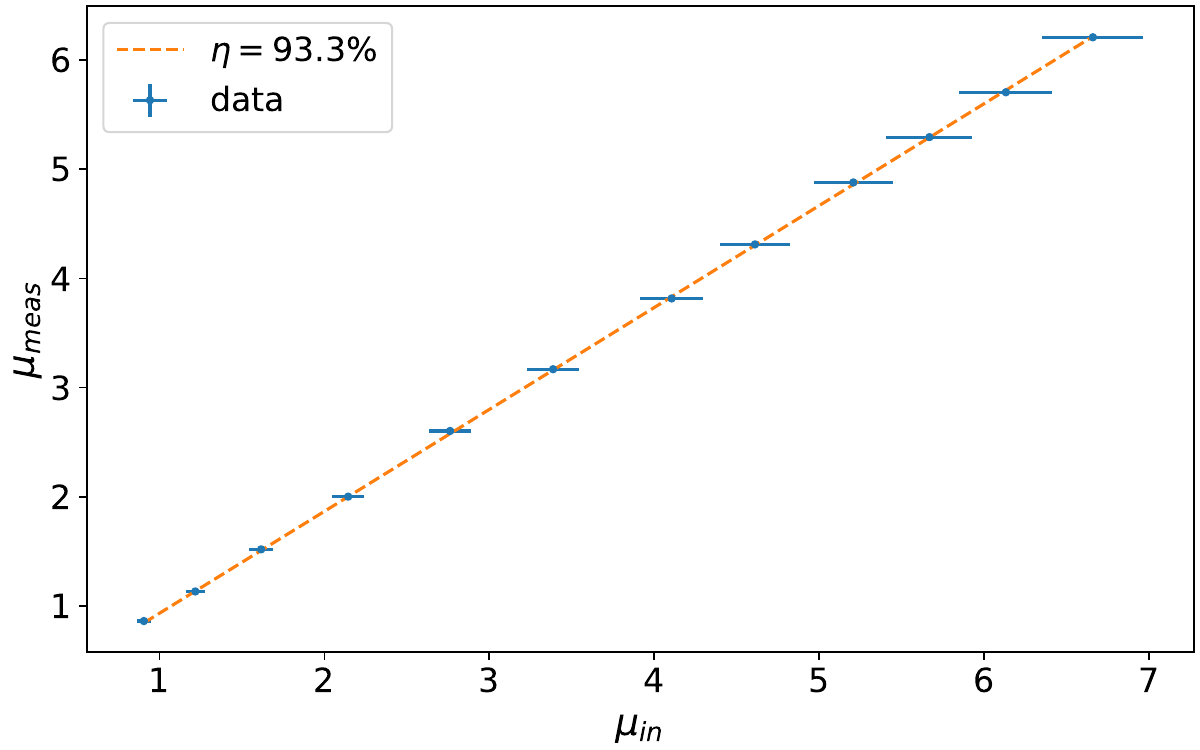}
    \caption{Calibration of the mean photon number per pulse measured by the TES using the IP method at 100~kHz, $\mu_{\text{meas}}$. The TES measurements are calibrated against an independent power meter, with values $\mu_{\text{in}}$. Data measured is recorded for 12 coherent states with varying amplitudes. The y-axis error bars are too small to be displayed. The fit to the data is performed by orthogonal distance regression estimating a 93.3$\pm$0.6\% detector efficiency.}
    \label{supp fig: detector efficiency}
\end{figure}

In the Methods section of the main article, we described how the amplitude of the coherent states for detector tomography can be measured. Figure \ref{supp fig: detector efficiency} plots the mean photon number measured by the IP method at 100~kHz against values measured by the independent Newport 918-IG-OD3R power meter. The power meter specification suggests a 2.6\% uncertainty in its readout, which also constitutes our estimate of calibration errors of the attenuators and beamsplitters. Combining calibration and reading errors, we estimate a 4.6\% overall uncertainty on the power meter's estimate of mean photon number, which is the y-axis error bar in Fig. \ref{supp fig: detector efficiency}. However, we did not account for the power meter non-linearity error, which is specified to be 0.5\% by the manufacturer, and the Fresnel reflection from the unterminated fibre plugged into the power meter, whose effect is to underestimate the photon number. 

The x-axis error bar in Fig. \ref{supp fig: detector efficiency}, which is the uncertainty on the mean photon number measured by the IP method at 100~kHz, is estimated by Poissonian counting statistics. A fibre mating coupler is used to connect the light source and the single-mode fibre leading to the TES, with a further splice inside the dilution fridge. We do not individually estimate the loss due to the mating coupler and the fibre splice, but absorb them into the overall system efficiency estimate. 

By applying orthogonal distance regression from the Scipy Python package~\cite{2020SciPy-NMeth} to the data shown in Fig. \ref{supp fig: detector efficiency}, the detector's efficiency is estimated to be 93.3$\pm$0.6\%. 

\section{Detector response to higher photon numbers at higher repetition rates}

\begin{figure}[!t]
    \centering
    \includegraphics[width=1.\textwidth]{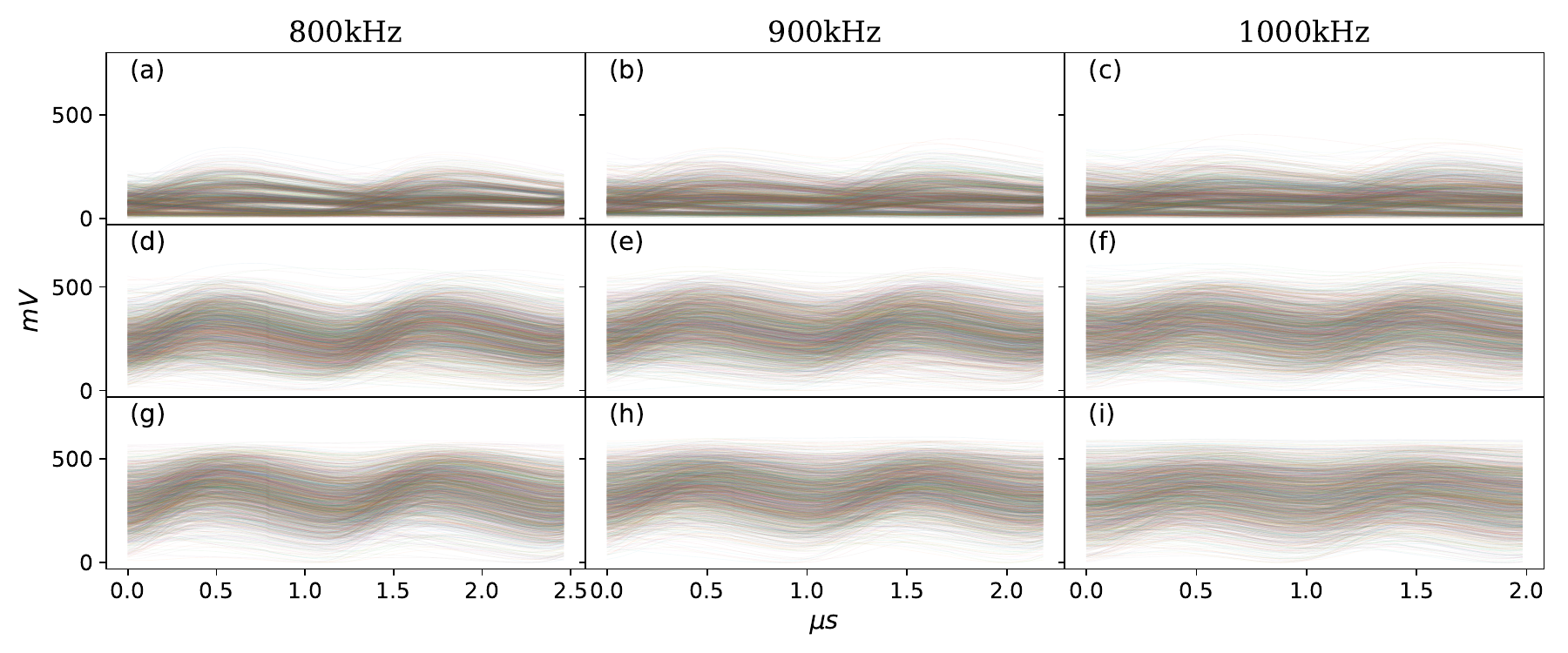}
    \caption{2000 raw TES voltage traces from coherent states at 800~kHz (left), 900~kHz (middle) and 1~MHz (right). The coherent state has mean photon number (a-c) $\mu=$0.86, (d-f) $\mu=$3.82, and (g-i) $\mu=$5.7.}
    \label{supp fig: raw traces}
\end{figure}

In the main article, we showed that the supervised learning methods have poor accuracy when detecting coherent states above 800~kHz and for higher photon number, as measured by the TVD of the reconstructed photon number distribution with the reference distribution. Figure \ref{supp fig: raw traces} shows that the voltage traces flatten at higher repetition rates and higher coherent state amplitudes, even though the voltage is far from saturating the voltage range of the analogue-to-digital converter, which is set to $\pm$1~V. The training step in supervised learning methods does not account for this change in signal shape, contributing to the algorithm's degraded performance. 

One possible reason for the flattening effect is due to the 1~MHz low-pass filter added to the room-temperature amplifier, which is used to filter out high frequency background noise but may have also removed certain attributes of the high-rate signal. Another likely reason is that the accumulated heat from consecutive absorptions of large photon numbers, without sufficient thermal dissipation, saturates the sensor causing the tungsten thin film to transition into normal conducting phase which temporarily diminishes its PNR capability. 

\end{document}